\documentclass[fleqn,showpacs,nofootinbib]{revtex4}

\usepackage{amssymb,amsmath,epsfig}

\DeclareMathOperator{\tr}{Tr}

\def\slash#1{\setbox0=\hbox{$#1$}               
        \dimen0=\wd0                            
        \setbox1=\hbox{/} \dimen1=\wd1          
        \ifdim\dimen0>\dimen1                   
        \rlap{\hbox to \dimen0{\hfil/\hfil}}    
        #1                                      
        \else                                   
        \rlap{\hbox to \dimen1{\hfil$#1$\hfil}} 
        /                                       
        \fi}                                    %

\begin{document}

\title{T-odd effects in photon-jet production at the Tevatron}

\author{D. Boer}
\email{D.Boer@few.vu.nl}
\affiliation{
Department of Physics and Astronomy, Vrije Universiteit Amsterdam,\\
NL-1081 HV Amsterdam, the Netherlands}

\author{P.J. Mulders}
\email{mulders@few.vu.nl}
\affiliation{
Department of Physics and Astronomy, Vrije Universiteit Amsterdam,\\
NL-1081 HV Amsterdam, the Netherlands}

\author{C. Pisano}
\email{cpisano@few.vu.nl}
\affiliation{
Department of Physics and Astronomy, Vrije Universiteit Amsterdam,\\
NL-1081 HV Amsterdam, the Netherlands}

\begin{abstract}
The angular distribution in photon-jet production in
$p \,  \bar{p} \rightarrow \gamma \,{\rm jet} \,X$ is
studied within a generalized factorization scheme
taking into account the transverse momentum of the partons in the initial
hadrons. Within this scheme an anomalously large $\cos 2\phi$ asymmetry
observed in the Drell-Yan process could be attributed to
the T-odd, spin and transverse momentum dependent
parton distribution function $h_1^{\perp\,q}(x, \boldsymbol p_{\perp}^2)$.
This same function is expected to produce a $\cos 2\phi$
asymmetry in the photon-jet production cross section.
We give the expression for
this particular azimuthal asymmetry, which is estimated to be smaller than the
Drell-Yan asymmetry but still of considerable size for Tevatron kinematics.
This offers a new possibility to study T-odd effects at the Tevatron.
\end{abstract}

\date{\today}

\maketitle
\section{Introduction}
It is well-known that the angular distribution of Drell-Yan lepton pairs
displays an anomalously large $\cos 2\phi $ asymmetry. This was
experimentally investigated using $\pi^-$ beams scattering off
deuterium and tungsten targets at center of mass energies of order 20 GeV
\cite{Falciano,Guanziroli,Conway}.
A next-to-leading order (NLO) analysis in perturbative QCD (pQCD) within the
standard framework of collinear factorization
failed to describe the data \cite{Brandenburg-93}.
More specifically, the observed violation of the so-called
Lam-Tung relation \cite{Lam-78,Lam-80,AL-82}, a relation between two angular
asymmetry terms, could not be described.
The NLO pQCD result is an order of magnitude
too small and of opposite sign. This has prompted much theoretical work
\cite{Brandenburg-93,Brandenburg-94,Eskola-94,Boer:1999mm,Boer:2002ju,Lu:2004hu,Boer-04,Lu:2005rq,Gamberg:2005ip,Brandenburg:2006xu},
offering explanations that go beyond the framework of collinear
factorization and/or leading twist perturbative QCD.

More recently, $p \, d$ Drell-Yan scattering was studied in a fixed target
experiment ($\sqrt{s} \approx 40$ GeV) at Fermilab \cite{Zhu:2006gx}.  The
angular distribution does not display a large $\cos 2\phi $ asymmetry,
indicating that the effect that causes the large asymmetry in $\pi^- \, N$
scattering is probably small for nonvalence partons. For this reason one would
like to investigate $p \, \bar{p}$ scattering, which is expected to be similar
to the $\pi^- \, N$ case (an expectation supported by model calculations
\cite{Boer:2002ju,Gamberg:2005ip,Barone:2006ws}).  It is an experiment that
could be done at the planned GSI-FAIR facility. It can in principle also be
done at Fermilab, although the energy of the collisions is so much higher
($\sqrt{s}=1.96 $ TeV) that the Drell-Yan asymmetry may be quite different in
magnitude, possibly much smaller at very high invariant mass $Q$ of the lepton
pair. Nevertheless, it would be interesting to see if NLO pQCD expectations
hold at those energies. Recently, such a study of the angular distribution
was done for $W$-boson production at the Tevatron and a nonzero result
compatible with NLO pQCD \cite{Mirkes:1992hu,Mirkes:1994dp} was obtained
\cite{Acosta:2005dn}.  This may likely be due to the fact that chirality flip
effects, such as the T-odd effect to be discussed here,
do not contribute to the $\cos 2\phi $ angular distribution
\cite{Bourrely:1994sc,Boer:2000er}. For neutral boson production they do
contribute however and therefore could lead to quite a different result. This
remains to be investigated.

In this paper we consider an asymmetry in the process $p\, \bar{p} \to \gamma \
\text{jet} \ X$ that potentially probes the same underlying mechanism and
could have certain advantages over Drell-Yan. This photon-jet
production process has already been studied experimentally in the
angular integrated case at the Tevatron
\cite{Kumar:2007mf}. Here we
will calculate the angular dependence within the framework as employed
in Ref.\ \cite{Boer:1999mm}, where transverse momentum and spin
dependence of partons inside hadrons is included\footnote{Photon-jet angular 
correlations in $p\, p$ and
$p\, \bar{p}$ collisions have also recently been studied in Ref.\
\cite{Pietrycki:2007xr} using the $k_t$-factorization approach applicable 
at small $x$, where gluon-gluon scattering dominates.}. 
In that case a
nontrivial polarization-dependent quark distribution (denoted by
$h_1^{\perp\,q}$) appears, which offers an explanation for the
anomalous angular asymmetry in the Drell-Yan process. The new
asymmetry is proportional to the analyzing power of the Drell-Yan
$\cos 2\phi $ asymmetry at the scale set by the transverse momentum of
the photon or the jet.  The latter asymmetry is expected to decrease
with increasing scale \cite{Boer:2001he}, but as we will demonstrate
the proportionality factor increases, leading one to expect a
significant asymmetry also at higher energies.

In Section II we discuss the theoretical framework and
the expected contributions to the new asymmetry. In Section III we study the
phenomenology of this asymmetry, using typical Tevatron kinematics and cuts.
We end with a summary of the results and the required measurement.
  
\section{Theoretical Framework: Calculation of the cross section}
We consider the process 
\begin{equation}
h_1(P_1){+}h_2(P_2)\, {\rightarrow}\,\gamma(K_\gamma){+}{\rm jet}(K_j){+}X \, ,
\end{equation}
where the four-momenta of the particles are given within brackets, and
the photon-jet pair in the final state is almost back-to-back in the plane perpendicular to the direction of the incoming hadrons. 
To lowest order in pQCD the reaction is described in terms of the partonic 
two-to-two subprocesses
\begin{equation}
q (p_1) + \bar{q}(p_2) \rightarrow \gamma (K_\gamma) + g(K_j)\, ,\qquad 
 {\rm and}\qquad
q (p_1) + {g}(p_2) \rightarrow \gamma (K_\gamma) + q(K_j)~.
\label{eq:sub}
\end{equation}
We make a lightcone decomposition of the hadronic momenta in terms of two light-like Sudakov vectors $n_+$ and $n_-$, satisfying $n_+^2\,{=}\,n_-^2\,{=}\,0$ and $n_+{\cdot}n_-\,{=}\,1$:
\begin{equation}
P_1^\mu
=P_1^+n_+^\mu+\frac{M_1^2}{2P_1^+}n_-^\mu\ ,\qquad\text{and}\qquad
P_2^\mu
=\frac{M_2^2}{2P_2^-}n_+^\mu+P_2^-n_-^\mu\ ~.
\end{equation}
In general $n_+$ and $n_-$ will define the lightcone components of every
vector $a$ as $a^\pm \equiv a \cdot n_\mp$, while 
perpendicular vectors $a_\perp$  will always refer to the components of  
 $a$ orthogonal to both incoming hadronic momenta, $P_1$ and $P_2$. Hence
the partonic momenta ($p_1$, $p_2$) can be expressed  in terms of the  
lightcone momentum fractions ($x_1$, $x_2$) and the 
intrinsic transverse momenta ($ p_{1 \perp}$, $ p_{2 \perp}$), as follows
\begin{equation}
p_1^\mu
=x_1^{\phantom{+}}\!P_1^+n_+^\mu
+\frac{m_1^2{+}\boldsymbol p_{1\perp}^2}{2x_1^{\phantom{+}}\!P_1^+}n_-^\mu
+p_{1\perp}^\mu\ ,
\qquad\text{and}\qquad
p_2^\mu
=\frac{m_2^2{+}\boldsymbol p_{2\perp}^2}{2x_2^{\phantom{-}}\!P_2^-}n_+^\mu
+x_2^{\phantom{-}}\!P_2^-n_-^\mu+p_{2\perp}^\mu\ .\label{PartonDecompositions}
\end{equation}
We denote with $s$ the total energy squared in the hadronic 
center-of-mass (c.m.) frame, $s =(P_1+P_2)^2 = E^2_{\rm c.m.} $, and 
 with $\eta_i$  the pseudo-rapidities 
of the outgoing particles,
\emph{i.e.}\ $\eta_i\,{=}\,{-}\ln\big(\tan(\frac{1}{2}\theta_i)\big)$,  $\theta_i$ being the polar angles of the outgoing particles in the same frame. 
Finally, we introduce the  partonic Mandelstam variables  
\begin{equation}
\hat s = (p_1 + p_2)^2, \qquad \hat t = (p_1-K_\gamma)^2 ,\qquad 
\hat u = (p_1-K_j)^2, 
\end{equation}
which satisfy the relations
\begin{equation}
\label{Yexpression}
 -\frac{\hat t}{\hat s} 
\equiv y = \frac{1}{e^{\eta_\gamma -\eta_j}\,{+}\,1}~ , \qquad {\rm and} \qquad   -\frac{\hat u}{\hat s} = 1-y~.
\end{equation}

\begin{figure}[t]
\epsfig{figure=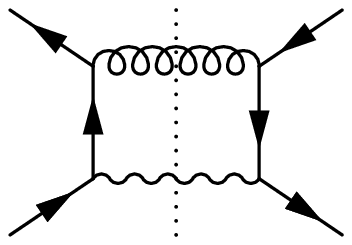,width=3cm}  \hspace{0.5cm}
\epsfig{figure=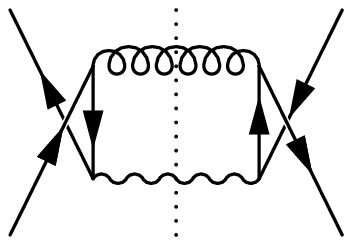,width=3cm}  \hspace{0.5cm}
\epsfig{figure=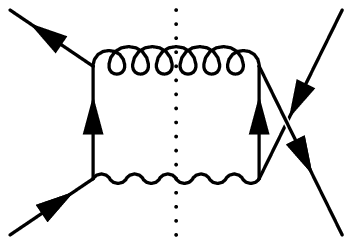,width=3cm}  \hspace{0.5cm}
\epsfig{figure=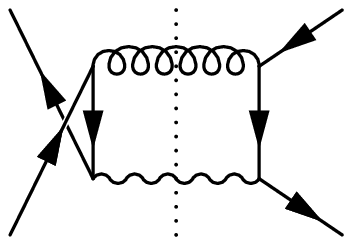,width=3cm}
\caption{Cut diagrams for the subprocess $q \bar q \rightarrow \gamma g$.}
\label{fig:qq}
\vspace{1cm}
\epsfig{figure=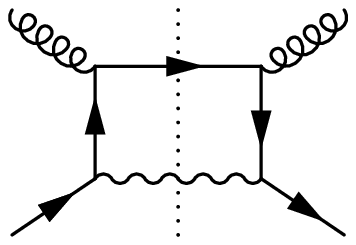,width=3cm}  \hspace{0.5cm}
\epsfig{figure=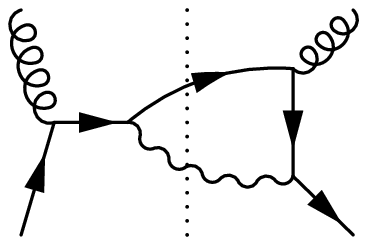,width=3cm}  \hspace{0.5cm}
\epsfig{figure=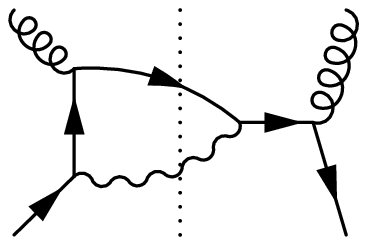,width=3cm}  \hspace{0.5cm}
\epsfig{figure=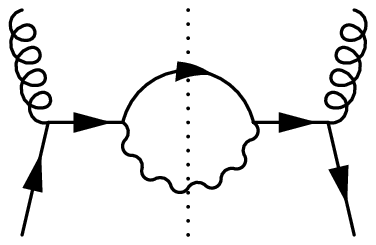,width=3cm}
\caption{Cut diagrams for the subprocess $q g\rightarrow \gamma q$.}
\label{fig:qg}
\end{figure}

Following Ref.\ \cite{Bacchetta:2007sz} 
we assume that at sufficiently high energies the 
hadronic cross section 
factorizes in a soft parton correlator for each observed hadron and a hard 
part:
\begin{eqnarray}
d\sigma^{h_1 h_2 \rightarrow \gamma {\rm jet} X}  
& = &\frac{1}{2 s}\, \frac{d^3 K_\gamma}{(2\pi)^3\,2E_\gamma}
\frac{d^3K_j}{(2\pi)^3\,2E_j}
{\int}d x_1\, d^2\boldsymbol p_{1\perp}\,d x_2\, d^2\boldsymbol p_{2\perp}\, (2\pi)^4
\delta^4(p_{1}{+}p_{2}{-}K_{\gamma}{-}K_{j})
 \nonumber \\
&&\qquad \qquad\qquad \qquad\qquad\times \sum_{a{,} b{,} c}\ 
\Phi_a(x_1{,}p_{1\perp})\otimes\Phi_b(x_2{,}p_{2\perp})
\otimes\,|H_{ab\rightarrow \gamma c}(p_1, p_2, K_{\gamma}, K_j)|^2\ ,
\label{CrossSec}
\end{eqnarray}
where the sum runs over all the incoming and outgoing partons
taking part in the subprocesses in (\ref{eq:sub}).
The convolutions $\otimes$ indicate the appropriate traces over Dirac 
indices and $|H|^2$ is the hard partonic squared amplitude,
obtained from the cut diagrams in Figs.\ \ref{fig:qq} and 
\ref{fig:qg} \cite{Bacchetta:2007sz}. 
The parton correlators are  defined on the lightfront LF ($\xi{\cdot}n\,{\equiv}\,0$, 
with $n\equiv n_-$ for parton 1 and $n\equiv n_+$ for parton 2); they describe 
the hadron $\rightarrow$ parton transitions and can be parameterized in terms 
of transverse momentum dependent (TMD) distribution functions. 
The quark content of an unpolarized hadron is described, 
in the lightcone gauge $A \cdot n=0$ and at leading twist, by the correlator 
\cite{Boer:1997nt}
\begin{eqnarray}
\label{QuarkCorr}
\Phi_q(x{,} p_\perp{;} P)
=  {\int}\frac{d(\xi{\cdot}P)\,d^2\xi_\perp}{(2\pi)^3}\ e^{ip\cdot\xi}\,
\langle P |\,\overline\psi(0)\,
\psi(\xi)\,|P\rangle\,\big\rfloor_{\text{LF}}
=  \frac{1}{2}\,
\bigg \{\,f_1^q(x{,}\boldsymbol{p}_\perp^2)\;\slash P
+i h_1^{\perp\,q}(x{,}\boldsymbol{p}_\perp^2)\;\frac{[\slash p_\perp , 
\slash P]}{2 M}\bigg \}\, , 
\end{eqnarray}
where $f_1^q(x, \boldsymbol{p}_\perp^2)$ is the unpolarized quark 
distribution, which integrated over $\boldsymbol{p}_\perp$ gives the familiar 
lightcone momentum distribution $f_1^q(x)$.  
The time-reversal (T) odd function  
$h_1^{\perp q}(x, \boldsymbol{p}_\perp^2)$ is interpreted as the quark 
transverse spin distribution in an unpolarized hadron
\cite{Boer:1997nt}. Below we will discuss the T-odd nature of this
function and its consequences in more detail. 

Analogously, for an antiquark,
\begin{eqnarray}
\label{AquarkCorr}
\bar\Phi_q(x{,} p_\perp{;} P)
=  -{\int}\frac{d(\xi{\cdot}P)\,d^2\xi_\perp}{(2\pi)^3}\ e^{-ip\cdot\xi}\,
\langle P |\,\overline\psi(0)\,
\psi(\xi)\,|P\rangle\,\big\rfloor_{\text{LF}}
=  \frac{1}{2}\,
\bigg \{\,f_1^{\bar q}(x{,}\boldsymbol{p}_\perp^2)\;\slash P
+i h_1^{\perp\,\bar q}(x{,}\boldsymbol{p}_\perp^2)\;\frac{[\slash p_\perp , 
\slash P]}{2 M}\bigg \}\, . 
\end{eqnarray}
The gluon correlator in the lightcone 
gauge is given by \cite{Mulders:2000sh}
\begin{eqnarray}
\label{GluonCorr}
\Phi_g^{\mu\nu}(x{,}p_\perp{;}P )
& =&  \frac{n_\rho\,n_\sigma}{(p{\cdot}n)^2}
{\int}\frac{d(\xi{\cdot}P)\,d^2\xi_\perp}{(2\pi)^3}\ e^{ip\cdot\xi}\,
\langle P|\,\tr\big[\,F^{\mu\rho}(0)\,
F^{\nu\sigma}(\xi)\,\big]
\,|P \rangle\,\big\rfloor_{\text{LF}} \nonumber \\
&=&\frac{1}{2x}\,\bigg \{-g_\perp^{\mu\nu}\,f_1^g(x{,}\boldsymbol{p}_\perp^2)
+\bigg(\frac{p_\perp^\mu p_\perp^\nu}{M^2}\,
{+}\,g_\perp^{\mu\nu}\frac{\boldsymbol p_\perp^2}{2M^2}\bigg)\;h_1^{\perp\,g}(x{,}\boldsymbol{p}_\perp^2) \bigg \}\, ,
\end{eqnarray}
with $g^{\mu\nu}_{\perp}$ being a transverse tensor defined  as
\begin{equation}
g^{\mu\nu}_{\perp} = g^{\mu\nu} - n_+^{\mu}n_-^{\nu}-n_-^{\mu}n_+^{\nu}\, .
\end{equation}
The function $f_1^g(x{,}\boldsymbol{p}_\perp^2)$ represents the usual 
unpolarized gluon distribution, while the T-even function $h_1^{\perp\,g}(x{,}\boldsymbol{p}_\perp^2)$ is the distribution of linearly polarized 
gluons in an unpolarized hadron. We include it here because it potentially
contributes to the observable of interest. However, it will turn out to yield 
a power-suppressed contribution. 


In the above expressions $h_1^{\perp\,q}$ appears as the T-odd part
in the parametrization of the correlator $\Phi_q$. For distribution 
functions, however, such a T-odd part is intimately connected to 
the gauge link that appears in the correlator $\Phi_q$ connecting 
the quark fields.
This gauge link is process-dependent and as a result, T-odd functions 
can appear with different factors in different processes. The specific 
factor can be traced back to the color flow in the cut diagrams of 
the partonic hard scattering. In situations in which only one
T-odd function contributes this has been
analyzed in detail for the process of interest and related processes, 
cf.\ e.g.\ Refs.\ \cite{Bomhof:2006ra,Bomhof:2007xt}. 
It leads to specific color factors multiplying the T-odd 
distribution function. In the case of single spin asymmetries (SSA)
one T-odd function appears, which could be  
the distribution function $h_1^{\perp\,q}$. Taking the appearance
in the case of leptoproduction as reference (having a factor $+1$ for 
the $\gamma^\ast q\rightarrow q$ subprocess), one finds appearance with
a different
factor in Drell-Yan scattering (factor $-1$ for $q\bar q\rightarrow 
\gamma^\ast$ subprocess \cite{Collins:2002kn,Belitsky:2002sm,Boer:2003cm})
or yet another factor for the appearance in the SSA in 
photon-jet production  
(factor $(N^2+1)/(N^2-1)$ for all contributions shown 
in Figs~\ref{fig:qq} and \ref{fig:qg} \cite{Bacchetta:2007sz}).
In the $\cos 2\phi$ asymmetry in Drell-Yan and in photon-jet production the
situation is different because a product of two T-odd functions arises
($h_1^{\perp q}\,h_1^{\perp\bar q}$).  
As we will discuss the factors are in both cases simply $+1$, which fixes the
relative sign of the asymmetries in the two processes to be $+1$.  
 
The factor in photon-jet production arises in the following way.
All 
diagrams have only two color flow contributions with relative strength 
$N^2$ and $-1$.
Color averaging in the initial state gives the usual color factors 
$(N^2-1)/N^2$ for the $q\bar q\rightarrow \gamma g$ process 
and $(N^2-1)/N(N^2-1) = 1/N$ for the $qg\rightarrow \gamma q$ process.
Including gauge links in the TMD correlators one schematically has
\begin{eqnarray}
q\bar q\rightarrow \gamma g:&\quad &
\frac{N^2}{N^2-1}\,\Phi_q^{[+(\Box^\dagger)]}\otimes
\overline\Phi_q^{[+^\dagger(\Box)]}\otimes\vert H\vert^2
\otimes\Delta_g^{[-][-^\dagger]}
-\frac{1}{N^2-1}\,\Phi_q^{[-]}\otimes
\overline\Phi_q^{[-^\dagger]}\otimes\vert H\vert^2
\otimes\Delta_g^{[-][-^\dagger]},
\label{gp-1}
\\
q g\rightarrow \gamma q:&\quad &
\frac{N^2}{N^2-1}\,\Phi_q^{[-(\Box)]}\otimes
\overline\Phi_g^{[+][-^\dagger]}\otimes\vert H\vert^2
\otimes\Delta_q^{[-^\dagger]}
-\frac{1}{N^2-1}\,\Phi_q^{[+]}\otimes
\overline\Phi_g^{[+][-^\dagger]}\otimes\vert H\vert^2
\otimes\Delta_q^{[-^\dagger]},
\label{gp-2}
\end{eqnarray}
where the correlators now include gauge links,
$\Phi^{[+]} \sim \langle \overline\psi(0)\,U^{[+]}\psi(\xi)\rangle$ or
$\Phi_g^{[+][-^\dagger]} \sim 
\langle F(0)U^{[+]}(0,\xi)F(\xi)U^{[-]\dagger}\rangle$, etc. The 
gauge links $U^{[\pm]}$ are the future and past-pointing ones in which
the path runs via lightcone $\pm \infty$, respectively. The $(\Box)$
contributions indicate the presence of a 
$\tfrac{1}{N}{\rm Tr}(U^{[+]}U^{[-]\dagger})$ term.
Following Ref.~\cite{Bomhof:2007xt}, the correlator $\Phi_q$
(and similarly $\overline\Phi_q$) can be split into two parts,
$\Phi_q^{[+(\Box^\dagger)]} = \Phi_q^{[+]} +\delta\Phi_q^{[+(\Box^\dagger)]}$,
where the 
non-universal part $\delta\Phi_q$ vanishes upon $p_T$-integration
or $p_T$-weighting.
In what follows we will omit these non-universal parts, but in principle
they could contribute (and potentially spoil factorization as recently
discussed in Refs~\cite{Collins:2007nk,Vogelsang:2007jk,Collins:2007jp})
 when one considers cross sections that are 
differential in measured transverse momenta. 
Ideally one should consider performing the
appropriate weighting to remove their possible contributions altogether.
We, however, keep the expressions for the cross sections differential  because
 the weighting is often difficult in experimental data analyses and
 the relation with the Drell-Yan expressions is more straightforward. After
  weighting one would be left only with the 
TMD functions 
$\Phi_q^{[\pm]}$ and $\overline\Phi_q^{[\pm^\dagger]}$, in which the 
gauge link determines the sign with which T-odd functions are multiplied,
giving in $q\bar q$ scattering for $h_1^{\perp q}$ a positive sign from 
$\Phi_q^{[+]}$ and a negative sign from $\Phi_q^{[-]}$.
If no factors arise from the other correlators one sees that in SSA
involving only one T-odd function $h_1^{\perp q}$, this function
appears in photon-jet production with the abovementioned
factor $(N^2+1)/(N^2-1)$ but otherwise the normal partonic cross section. 
This factor and similar ones for other T-odd functions were used
in Ref.~\cite{Bacchetta:2007sz}. 
In the present case of a product of two T-odd functions, we 
obtain from both combinations of correlators
$\Phi_q^{[+]}\otimes \overline\Phi_q^{[+^\dagger]}$
and $\Phi_q^{[-]}\otimes\overline\Phi_q^{[-^\dagger]}$
in Eq.~(\ref{gp-1}) now a positive sign. Hence, the 
contribution $h_1^{\perp q}\,h_1^{\perp\bar q}$ appears with 
just a factor $+1$, justifying the use of the parametrizations
in Eqs~(\ref{QuarkCorr}) and (\ref{AquarkCorr}) in combination with the
normal partonic hard scattering amplitudes squared.
For the contribution involving the T-even gluon function
$h_1^{\perp g}$ no process dependent color factors need to be considered.

The issue of factorization is left as an open question and the same 
applies to
resummation. In Refs~\cite{Boer:2006eq,Berger:2007jw} the resummation for the
 $\cos 2\phi$ asymmetry 
is addressed and found to be unclear beyond the leading logarithmic
approximation. A similar situation could apply to the asymmetry in 
the photon-jet production case, but we will not address this here. Our present
goal is to point out how the contribution of $h_1^{\perp q}\,h_1^{\perp\bar q}$
enters the asymmetry expression in leading order and to give an estimate of 
its expected magnitude.


In order to derive an expression for the cross section in terms of parton 
distributions, we insert the parametrizations 
(\ref{QuarkCorr})-(\ref{GluonCorr}) of the TMD quark, antiquark and 
gluon correlators into (\ref{CrossSec}). Furthermore,    
utilizing the decompositions of the parton momenta in \eqref{PartonDecompositions},  the $\delta$-function in 
(\ref{CrossSec}) can be rewritten as
\begin{eqnarray}
\label{DeltaFunc}
\delta^4(p_1{+}p_2{-}k_1{-}k_2)
& = &
\frac{2}{s}\,
\delta\bigg(\,x_1{-}\frac{1}{\sqrt s}
(\,|\boldsymbol K_{\gamma\perp}|\,e^{\eta_\gamma}\,
{+}|\boldsymbol K_{j\perp}|\,e^{\eta_j}\,)\,\bigg)\,
\delta\bigg(\,x_2{-}\frac{1}{\sqrt s}
(\,|\boldsymbol K_{\gamma\perp}|e^{-\eta_\gamma}\,
{+}|\boldsymbol K_{j\perp}|\,e^{-\eta_j}\,)\,\bigg) \nonumber \\
&&\mspace{200mu}
\times\delta^2(\boldsymbol p_{1\perp}{+}\boldsymbol p_{2\perp}{-}\boldsymbol K_{\gamma\perp}{-}\boldsymbol K_{j\perp})\ ,
\end{eqnarray}
with corrections of order $\mathcal O(1/s^2)$.  After integration 
over $x_1$ and $x_2$, which fixes the parton momentum fractions by the first 
two $\delta$-functions on the r.h.s.~of  (\ref{DeltaFunc}), the resulting 
hadronic cross section consists of two contributions, 
{\it i.e.}
\begin{equation}
\frac{d \sigma^{h_1 h_2\to \gamma {\rm jet} X}}
{d\eta_\gamma\, d^2   \boldsymbol K_{\gamma\perp}  \, d\eta_j \,d^2 \boldsymbol K_{j\perp}\, d^2 \boldsymbol q_{\perp} } = 
\frac{d \sigma [f_1^{q, g}]}
{d\eta_\gamma\, d^2   \boldsymbol K_{\gamma\perp}  \, d\eta_j \,d^2 \boldsymbol K_{j\perp}\, d^2 \boldsymbol q_{\perp} } + \frac{d \sigma[h_1^{\perp\, q}]}
{d\eta_\gamma\, d^2   \boldsymbol K_{\gamma\perp}  \, d\eta_j \,d^2 \boldsymbol K_{j\perp}\, d^2 \boldsymbol q_{\perp} } \, ,
\label{eq:cso} 
\end{equation}
with $\boldsymbol{q}_\perp \equiv \boldsymbol{K}_{\gamma \perp} + \boldsymbol{K}_{j \perp}$. We are interested in events in which the photon and jet
are approximately back-to-back in the transverse plane, therefore  
$|\boldsymbol{q}_\perp| \ll |\boldsymbol{K}_{\gamma\perp}|, 
|\boldsymbol{K}_{j\perp}|$.  
The cross section $d\sigma[f_1^{q, g}]$ depends on
 the unpolarized (anti)quark and  gluon distributions $f_1^{q, g}$, while
 $d\sigma[h_1^{\perp\, q}]$ depends on the (anti)quark 
function $h_1^{\perp\, q}$. More explicitly, 
\begin{eqnarray}
\frac{d \sigma[f_1^{q, g}]}
{d\eta_\gamma\, d^2   \boldsymbol K_{\gamma\perp}  \, d\eta_j \,d^2 \boldsymbol K_{j\perp}\, d^2 \boldsymbol q_{\perp} }
&  =   & \frac{ \alpha \alpha_s} {s   \boldsymbol K_{\gamma\perp} ^2 }\,  
\delta^2 (\boldsymbol{q}_\perp - \boldsymbol K_{\gamma\perp} -\boldsymbol K_{j \perp})
\sum_q e_q^2  \int d^2\boldsymbol{p}_{1\perp}\,d^2
\boldsymbol{p}_{2\perp} 
\delta^2 (\boldsymbol{p}_{1\perp} +\boldsymbol{p}_{2\perp} -\boldsymbol{q}_{\perp}) \nonumber \\
& & \quad \times \bigg \{ \frac{1}{N}\,(1-y) (1+y^2)  \,
f_1^q(x_1, \boldsymbol{p}_{1\perp}^2) f_1^g(x_2, \boldsymbol{p}_{2\perp}^2)
\nonumber \\
&& \quad \quad + \frac{1}{N}\, y (1+(1-y)^2) f_1^q(x_2, \boldsymbol{p}_{2\perp}^2) f_1^g(x_1, \boldsymbol{p}_{1\perp}^2)  + \frac{N^2-1}{N^2}  \bigg [ (y^2 +(1-y)^2) 
\nonumber \\
&&\quad \quad \quad \times (f_1^q(x_1, \boldsymbol{p}_{1\perp}^2)f_1^{\bar q}(x_2, \boldsymbol{p}_{2\perp}^2) + f_1^q(x_2, \boldsymbol{p}_{2\perp}^2)f_1^{\bar q}(x_1, \boldsymbol{p}_{1\perp}^2)) \bigg ] \bigg \}\, , 
\label{eq:cs}
\end{eqnarray}
and
\begin{eqnarray}
\frac{d\sigma [h_1^{\perp\,q}]}
{d\eta_\gamma\, d^2   \boldsymbol K_{\gamma\perp}  \, d\eta_j \,d^2 \boldsymbol K_{j\perp}\, d^2 \boldsymbol q_{\perp} }
&  = &- \frac{2  \alpha \alpha_s} { s   \boldsymbol K_{\gamma\perp} ^2 }\,  
\delta^2 (\boldsymbol{q}_\perp - \boldsymbol K_{\gamma\perp} 
-\boldsymbol K_{j \perp})
\sum_q e_q^2  \int d^2\boldsymbol{p}_{1\perp}\,d^2
\boldsymbol{p}_{2\perp} 
\delta^2 (\boldsymbol{p}_{1\perp} +\boldsymbol{p}_{2\perp} -\boldsymbol{q}_{\perp}) \nonumber \\
&&\times \frac{y (1-y)}{M_1 M_2} \bigg ((\boldsymbol{p}_{1\perp}\cdot\boldsymbol{p}_{2\perp}) + \frac{ (\boldsymbol{K}_{\gamma\perp}\cdot \boldsymbol {p}_{1\perp}) 
(\boldsymbol{K}_{j\perp} \cdot \boldsymbol{p}_{2\perp}) +(\boldsymbol{K}_{\gamma\perp}\cdot \boldsymbol {p}_{2\perp}) 
(\boldsymbol{K}_{j\perp} \cdot \boldsymbol{p}_{1\perp}) }{ \boldsymbol{K}_{\gamma\perp} ^2} \bigg )\nonumber \\
&& \quad \quad\times  \frac{N^2-1}{N^2} ( h_1^{\perp \,q}(x_1, \boldsymbol{p}^2_{1\perp}) h_1^{\perp \,\bar q} 
(x_2, \boldsymbol{p}^2_{2\perp}) +  h_1^{\perp \,q}(x_2, \boldsymbol{p}^2_{2\perp}) h_1^{\perp \,\bar q} 
(x_1, \boldsymbol{p}^2_{1\perp}) ) \bigg \}\, , 
\label{eq:csphi}
\end{eqnarray} 
where the sums always run over quarks and antiquarks. 
Power-suppressed terms 
of the order ${\cal{O}}(1/(\boldsymbol K_{\gamma\perp}^4 s))$, such as the ones
proportional to the gluon distribution function  $h_1^{\perp\, g}$,   
are neglected throughout this paper. If we define the function
\begin{eqnarray}
{\cal{H}}(x_1, x_2, \boldsymbol{q}_\perp^2 ) & \equiv & \frac{1}{M_1 M_2}\sum_q e_q^2  \int d^2\boldsymbol{p}_{1\perp}\,d^2
\boldsymbol{p}_{2\perp} 
\delta^2 (\boldsymbol{p}_{1\perp} +\boldsymbol{p}_{2\perp} -\boldsymbol{q}_{\perp}) \bigg (2 (\hat{\boldsymbol{h}}_{\perp} \cdot \boldsymbol{p}_{1 \perp})( \hat{\boldsymbol{h}}_{\perp} \cdot \boldsymbol{p}_{2 \perp}) - (\boldsymbol{p}_{1 \perp} \cdot\boldsymbol{p}_{2 \perp}) \bigg )
\label{eq:H}
\nonumber \\
&& \qquad \times ( h_1^{\perp q}(x_1, \boldsymbol p^2_{1\perp}) h_1^{\perp \bar q} 
(x_2, \boldsymbol p^2_{2\perp}) +  h_1^{\perp q}(x_2, \boldsymbol p^2_{2\perp}) h_1^{\perp \bar q} 
(x_1, \boldsymbol p^2_{1\perp}) )\, ,
\end{eqnarray}
with $\hat{\boldsymbol{h}} \equiv \boldsymbol{q}_{\perp}/|\boldsymbol{q}_{\perp}|$, and denote with $\phi_\gamma$, $\phi_j$ and $\phi_\perp$ the azimuthal angles,
in the hadronic center-of-mass frame,
of the outgoing photon, jet and  vector $\boldsymbol{q}_\perp$ respectively,
then (\ref{eq:csphi}) can be rewritten as  
\begin{eqnarray}
\frac{d \sigma[h_1^{\perp\, q}]}
{d\eta_\gamma\, d^2   \boldsymbol K_{\gamma\perp}  \, d\eta_j \,d^2 \boldsymbol K_{j\perp}\, d^2 \boldsymbol q_{\perp} }
&  =   &  
-\frac{2 \alpha \alpha_s} {s   \boldsymbol K_{\gamma\perp} ^2 }\,
{y (1-y)}\, \frac{N^2-1}{N^2}
\delta^2 (\boldsymbol{q}_\perp - \boldsymbol K_{\gamma \perp} -\boldsymbol K_{j \perp}) \nonumber \\
&& \quad \quad  \times \,\sum_{l, m =1}^2\,
 \frac{\boldsymbol{K}_{\gamma \perp}^{\{ l}\boldsymbol{K}_{j \perp}^{m \}}}{2  \,\boldsymbol{K}_{\gamma \perp}^2 }
\bigg (\frac{\boldsymbol{q}_{\perp}^{\{ l}\boldsymbol{q}_{\perp}^{m\}}}{\boldsymbol{q}_{\perp}^2} - \delta^{l m} \bigg ){\cal{H}}(x_1, x_2, \boldsymbol{q}_\perp^2 )  \nonumber \\
& = & -\frac{2 \alpha \alpha_s} {s   \boldsymbol K_{\gamma\perp} ^2 }\,
{y (1-y)}\, \frac{N^2-1}{N^2}\, 
 \delta^2 (\boldsymbol{q}_\perp - \boldsymbol K_{\gamma \perp} -\boldsymbol K_{j \perp}) \nonumber \\
&& \quad \quad \quad \times \,\cos (2 \phi_\perp -\phi_\gamma -\phi_j) {\cal{H}}(x_1, x_2, \boldsymbol{q}_\perp^2 ).
\label{eq:csphi2}
\end{eqnarray}
The approximation $|\boldsymbol{K}_{\gamma \perp}| \approx |\boldsymbol{K}_{j \perp}|$ has been used in the derivation of the second equation in 
(\ref{eq:csphi2}).  Furthermore, the same  approximation allows us 
to derive the following relations, starting from the definition of 
$\boldsymbol{q}_\perp$ in terms of $\boldsymbol K_{\gamma\perp}$ and $\boldsymbol K_{j\perp}$, 
\begin{eqnarray}
\frac{|\boldsymbol{q}_\perp |}{|\boldsymbol{K}_{\gamma\perp}|}
 \cos\phi_\perp & = &   \cos\phi_\gamma + \cos\phi_j    \, ,    \nonumber \\ 
\frac{|\boldsymbol{q}_\perp|}{|\boldsymbol{K}_{\gamma\perp}|} \sin\phi_\perp & = & \sin\phi_\gamma + \sin\phi_j ~.
\label{eq:angles}
\end{eqnarray}
Since $|\boldsymbol{q}_\perp | \ll |\boldsymbol K_{\gamma\perp}|$, 
(\ref{eq:angles}) implies  $\cos\phi_j\approx -\cos\phi_\gamma$ and 
$\sin\phi_j\approx -\sin\phi_\gamma$, so that 
\begin{eqnarray}
\cos (2 \phi_\perp - \phi_\gamma -\phi_j) 
\approx  -\cos 2 (\phi_\perp -\phi_\gamma)
\approx  -\cos 2 (\phi_\perp -\phi_j)\, , 
\end{eqnarray} 
and (\ref{eq:csphi2}) takes the form
\begin{eqnarray}
\frac{d \sigma[h_1^{\perp\, q}]}
{d\eta_\gamma\, d^2   \boldsymbol K_{\gamma\perp}  \, d\eta_j \,d^2 \boldsymbol K_{j\perp}\, d^2 \boldsymbol q_{\perp} }
& = & \frac{2 \alpha \alpha_s} {s   \boldsymbol K_{\gamma\perp} ^2 }\,
{y (1-y)}\, \frac{N^2-1}{N^2}\,
  \delta^2 (\boldsymbol{q}_\perp - \boldsymbol K_{\gamma \perp} -\boldsymbol K_{j \perp})\cos 2(\phi_\perp  -\phi_j) {\cal{H}}(x_1, x_2, \boldsymbol{q}_\perp^2 )~.\nonumber \\
\label{eq:csphi3}
\end{eqnarray}
Substituting (\ref{eq:cs}) and (\ref{eq:csphi3}) into (\ref{eq:cso}), 
and integrating
over $\boldsymbol{K}_{j\perp}$ (alternatively, integrating over 
$\boldsymbol{K}_{\gamma\perp}$ would lead to the same equations as presented
below, but with the 
replacement $\boldsymbol{K}_{\gamma\perp} \leftrightarrow 
\boldsymbol{K}_{j\perp}$), we obtain
\begin{eqnarray}
\frac{d\sigma^{h_1 h_2\to \gamma {\rm jet} X}}
{d\eta_\gamma\, d\eta_j \,d^2 \boldsymbol K_{\gamma\perp}\, 
d^2 \boldsymbol q_{\perp} } & = & \int d^2   \boldsymbol K_{j\perp} \,
\frac{d\sigma^{h_1 h_2\to \gamma {\rm jet} X}}
{d\eta_\gamma\, d^2   \boldsymbol K_{\gamma\perp}  \, d\eta_j \,d^2 \boldsymbol K_{j\perp}\, d^2 \boldsymbol q_{\perp} }\nonumber \\
&  =   & \frac{ \alpha \alpha_s}{s   \boldsymbol K_{\gamma \perp} ^2 }\, \bigg \{ \frac{1}{N}\,(1-y) (1+y^2)  \,{\cal{G}}(x_1, x_2, \boldsymbol{q}_{\perp}^2)+ \frac{1}{N}\, y (1+(1-y)^2)\, {\cal{\tilde{G}}}(x_1, x_2, \boldsymbol{q}_{\perp}^2) \nonumber \\
&& + \frac{N^2-1}{N^2}  \bigg [ (y^2+(1-y)^2) {\cal{F}}(x_1,x_2,\boldsymbol{q}_\perp^2 )  +  \, 2 y (1-y)\, {{\cal{H}}}(x_1, x_2, \boldsymbol{q}_\perp^2 ) \,\cos 2(\phi_\perp  -\phi_\gamma)\bigg ] \bigg \}\,
\nonumber \\ 
\label{eq:cross}
\end{eqnarray}
where, in analogy to (\ref{eq:H}), the following convolutions of distribution 
functions have been utilized:
\begin{eqnarray}
{\cal{F}}(x_1, x_2, \boldsymbol{q}_{\perp}^2) & \equiv & \sum_q e_q^2  \int d^2\boldsymbol{p}_{1\perp}\,d^2
\boldsymbol{p}_{2\perp} 
\delta^2 (\boldsymbol{p}_{1\perp} +\boldsymbol{p}_{2\perp} -\boldsymbol{q}_{\perp})  ( f_1^{q}(x_1, \boldsymbol p^2_{1\perp}) f_1^{\bar q} 
(x_2, \boldsymbol p^2_{2\perp}) +  f_1^{q}(x_2, \boldsymbol p^2_{2\perp}) 
f_1^{\bar q} 
(x_1, \boldsymbol p^2_{1\perp}) )\, , \nonumber \\
{\cal{G}}(x_1, x_2, \boldsymbol{q}_{\perp}^2) & \equiv & \sum_q e_q^2  \int d^2\boldsymbol{p}_{1\perp}\,d^2
\boldsymbol{p}_{2\perp} 
\delta^2 (\boldsymbol{p}_{1\perp} +\boldsymbol{p}_{2\perp} -\boldsymbol{q}_{\perp})   f_1^{ q}(x_1, \boldsymbol p^2_{1\perp}) f_1^g(x_2, \boldsymbol p^2_{2\perp})\, ,  \nonumber \\
\tilde{{\cal{G}}}(x_1, x_2, \boldsymbol{q}_{\perp}^2) & \equiv & \sum_q e_q^2  \int d^2\boldsymbol{p}_{1\perp}\,d^2
\boldsymbol{p}_{2\perp} 
\delta^2 (\boldsymbol{p}_{1\perp} +\boldsymbol{p}_{2\perp} -\boldsymbol{q}_{\perp})   f_1^{ q}(x_2, \boldsymbol p^2_{2\perp}) f_1^g(x_1, \boldsymbol p^2_{1\perp})\, .
\label{def:conv}
\end{eqnarray}
Alternatively, equation (\ref{eq:cross}) can be rewritten as 
\begin{eqnarray}
\frac{d\sigma^{h_1 h_2\to \gamma {\rm jet} X}}
{d\eta_\gamma\, d\eta_j \,d^2 \boldsymbol K_{\gamma\perp}\, d^2 \boldsymbol q_{\perp} }   
&  =  & \frac{1}{\pi^2}\,\frac{d\sigma^{h_1 h_2\to \gamma {\rm jet} X}}
{d\eta_\gamma\, d\eta_j \,d \boldsymbol K_{\gamma\perp}^2\, d \boldsymbol q_{\perp}^2 } \,   \bigg (1 + {\cal{A}}(y, x_1, x_2, \boldsymbol{q}_\perp^2) \, 
\cos 2 (\phi_\perp -\phi_\gamma)    \bigg )\, ,
\label{eq:csfinal} 
\end{eqnarray}
with
\begin{eqnarray}
\frac{d \sigma^{h_1 h_2\to \gamma {\rm jet} X}}
{d\eta_\gamma\, d\eta_j \,d \boldsymbol K_{\gamma\perp}^2\, 
d \boldsymbol q_{\perp}^2 } & = & \frac{1}{4}\int d\phi_\perp\, d\phi_\gamma \, d^2   \boldsymbol K_{j\perp} \,
\frac{d \sigma^{h_1 h_2\to \gamma {\rm jet} X}}
{d\eta_\gamma\, d^2   \boldsymbol K_{\gamma\perp}  \, d\eta_j \,d^2 \boldsymbol K_{j\perp}\, d^2 \boldsymbol q_{\perp} }\nonumber \\
&  =   & \frac{\pi^2 \alpha \alpha_s}{s   \boldsymbol K_{\gamma \perp} ^2 }\, \bigg \{ \frac{1}{N}\,(1-y) (1+y^2)  \,{\cal{G}}(x_1, x_2, \boldsymbol{q}_{\perp}^2)+ \frac{1}{N}\, y (1+(1- y)^2)\, {\cal{\tilde{G}}}(x_1, x_2, \boldsymbol{q}_{\perp}^2) \nonumber \\
&& + \frac{N^2-1}{N^2} \,  (y^2+(1-y)^2) {\cal{F}}(x_1,x_2,\boldsymbol{q}_\perp^2 )  \bigg \}
\label{eq:cstotal}
\end{eqnarray}
and the azimuthal asymmetry 
\begin{eqnarray}
{\cal{A}}(y, x_1, x_2, \boldsymbol{q}_\perp^2) & = & 
 \nu (x_1, x_2, \boldsymbol{q}_\perp^2) \,{R}(y, x_1, x_2, \boldsymbol{q}_\perp^2) \, , 
\label{eq:asymmetry}
\end{eqnarray}
where
\begin{eqnarray}
\nu(x_1, x_2,\boldsymbol{q}_{\perp}^2 ) &  = & \frac{2\, {\cal{H}}(x_1, x_2, \boldsymbol{q}_\perp^2)}{{\cal{F}} (x_1, x_2, \boldsymbol{q}_\perp^2)}\,  
\label{eq:nu}
\end{eqnarray}
contains the dependence on $h_1^{\perp\,q}$ and is identical to the azimuthal
asymmetry expression that appears in the Drell-Yan process \cite{Boer:1999mm}, 
with the scale $Q$ equal to $|\boldsymbol{K}_{\gamma \perp}|$. 
The ratio
\begin{eqnarray}
{R}& = &  \frac{\pi^2\alpha\alpha_s}{s \boldsymbol{K}_{\gamma \perp}^2}\, y (1-y)\, {\cal{F}} (x_1, x_2, \boldsymbol{q}_\perp^2)\, \bigg (\frac{d\sigma^{h_1 h_2\to \gamma {\rm jet} X}}
{d\eta_\gamma\, d\eta_j \,d \boldsymbol K_{\gamma\perp}^2\, d \boldsymbol q_{\perp}^2 } \bigg )^{-1} \nonumber \\
& = & 
\frac{{N^2 \, y (1-y)\,\cal{F}} (x_1, x_2, \boldsymbol{q}_\perp^2)}{N \, (1-y)
  (1+y^2)  \,{\cal{G}}(x_1, x_2, \boldsymbol{q}_{\perp}^2)+ N \, 
y (1+(1-y)^2)\, {\cal{\tilde{G}}}(x_1, x_2, \boldsymbol{q}_{\perp}^2)  + (N^2-1) \, (y^2+(1-y)^2) {\cal{F}}(x_1,x_2,\boldsymbol{q}_\perp^2 )}\nonumber \\
\label{eq:ratio}
\end{eqnarray}
only depends on the T-even distribution functions $f_1^{q, g} (x, \boldsymbol{p}           _{\perp}^2)$. 

\section{Phenomenology: the azimuthal asymmetry}

\begin{figure}[t]
\begin{center}
\epsfig{figure=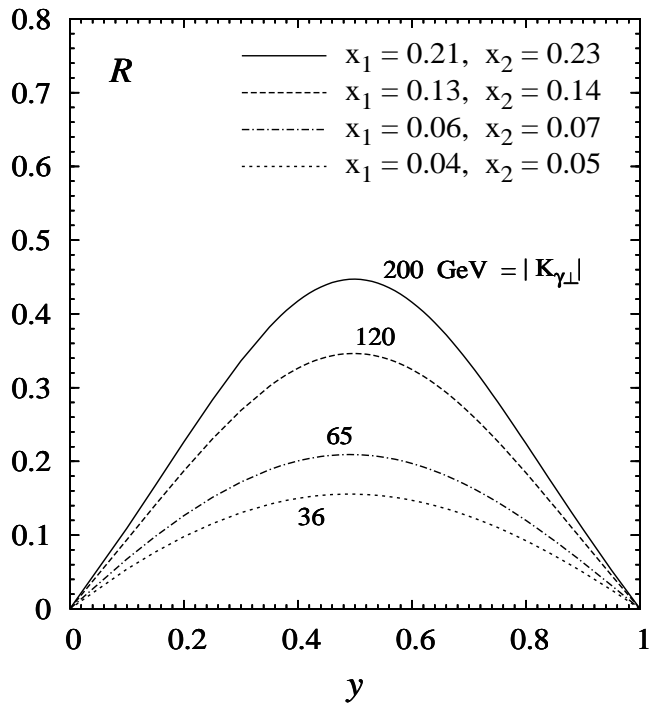, width = 10cm, height = 9.5cm}
\caption{The ratio $R$ defined in (\ref{eq:ratio}) as a function of $y$, 
calculated according to (\ref{eq:ratio2}) for different values of
$x_1$, $x_2$, $|\boldsymbol{K}_{\gamma\perp}|$ typical of the Tevatron 
experiments \cite{D0:Atr}.} 
\label{fig:ratio}
\end{center}
\end{figure}
The process $p \,\bar{p} \rightarrow \gamma \,{\rm jet}\, X $ is currently 
being analyzed by the D\O\ Collaboration at the  Tevatron  collider 
\cite{Kumar:2007mf,D0:Atr}. 
Data on the cross section, differential in 
$\eta_\gamma$, $\eta_j$ and $\boldsymbol{K}_{j\perp}^2$, have been taken at 
$\sqrt{s} = 1.96$ TeV, and were considered in a preliminary study
\cite{D0:Atr} with the following kinematic cuts:
\begin{eqnarray}
&&|\boldsymbol K_{\gamma\perp}| > 30 ~{\rm GeV}\, ,\qquad  -1 <\eta_\gamma < 1 
~~({\rm central~ region}) \nonumber \\
&&|\boldsymbol K_{j\perp}| > 15 ~{\rm GeV}\, ,\qquad  -0.8 \le \eta_j \le 0.8 ~~({\rm central}) \, , \qquad 1.5 <|\eta_j| < 2.5 ~~({\rm forward});
\end{eqnarray}
the transverse momentum imbalance between the photon and the jet being 
constrained by the relation
\begin{equation}
|\boldsymbol{q}_\perp|  < 12.5 + 0.36 \times 
|\boldsymbol{K}_{\gamma\perp}|~(\rm {GeV})~.
\end{equation}
Such angular integrated measurements are only sensitive to the 
transverse momentum integrated parton distributions
\begin{equation}
f_1^{q, g}(x) = \int d \boldsymbol p_{\perp}^2\, f_1^{q, g}(x, \boldsymbol p_{\perp}^2)~,
\end{equation}
as can be seen from the leading order expression of the cross section, 
\begin{eqnarray}
\frac{d\sigma^{p \bar{p} \rightarrow \gamma {\rm jet} X}}
{d\eta_\gamma\, d\eta_j\, d \boldsymbol{K}^2_{\gamma \perp}} & = & \frac{ \pi \alpha \alpha_s} {s   \boldsymbol K_{\gamma\perp} ^2 }
\sum_q e_q^2   \bigg \{ \frac{1}{N}\,(1-y) (1+y^2)  \,
f_1^q(x_1) f_1^g(x_2) + \frac{1}{N}\, y (1+(1-y)^2) f_1^q(x_2) f_1^g(x_1)
\nonumber \\
&&\qquad  \qquad+ 2 \frac{N^2-1}{N^2}(y^2+(1-y)^2) 
f_1^q(x_1){f}_1^{q}(x_2)\bigg \}\, , 
\end{eqnarray}
obtained after integrating (\ref{eq:cstotal}), together with the definitions in
(\ref{def:conv}), over $\boldsymbol{q}^2_\perp$. Here we have used that the 
antiquark contribution in the antiproton equals the quark
contribution inside a proton.  
A study of the angular dependent 
cross section in (\ref{eq:csfinal}) will provide valuable information on the 
TMD distribution function $h_1^{\perp\,q}(x, \boldsymbol{p}_\perp^2)$, if the
azimuthal asymmetry ${\cal{A}}$ turns out to be sufficiently 
sizeable in the available kinematic region. 
Model calculations \cite{Boer:2002ju,Barone:2006ws} 
applied to the $p\,\bar p$ Drell-Yan process
have shown that the quantity $\nu$ in (\ref{eq:asymmetry})
 is of the order of 30\% or higher for $|\boldsymbol{q}_\perp|$ of a few GeV
 and $Q$ values of ${\cal O}(1-10)$ GeV. 
Therefore, a study of the order of magnitude of ${\cal A}$ 
as a function of $x_1$, $x_2$ and $\boldsymbol{q}_\perp^2$ requires 
an estimate of the ratio $R$ defined in (\ref{eq:ratio}). 
This will be obtained as follows. 

First of all, the  
unknown TMD distribution functions appearing in (\ref{eq:ratio}) are 
 evaluated assuming a
factorization of their transverse momentum dependence (see, for example, 
\cite{Mulders:1995dh,D'Alesio:2004up}), that is  
\begin{equation}
f_1^{q, g}(x_, \boldsymbol{p}_{\perp}^2) = f_1^{q, g}(x) 
{\cal T}(\boldsymbol{p}_\perp^2)~,
\end{equation}
with  $f_1^{q, g}(x)$ being the usual unpolarized parton distributions and 
${\cal T}(\boldsymbol{p}_\perp^2)$ being a generic function, taken to be the 
same for all partons and often chosen to be Gaussian. The
$\boldsymbol{q}_\perp^2$-dependence of $R$ then drops out and (\ref{eq:ratio})
takes the form  
\begin{equation}
R = \frac{2 N^2 y (1-y) \sum_q e^2_q \,f_1^q(x_1)f_1^q(x_2)}{\sum_q e^2_q \big \{{N} (1-y)(1+y^2)f_1^q(x_1)f_1^g(x_2) + {N} y (1+(1-y)^2)f_1^q(x_2)f_1^g(x_1) + 2\, (N^2-1)(y^2+(1-y)^2)  f_1^q(x_1)f_1^q(x_2) \big \} }~.
\label{eq:ratio2}
\end{equation}
We consider only light quarks, {\it i.e.\/} the sum in (\ref{eq:ratio2})   
runs over 
$q = u$, $\bar{u}$, $d$, $\bar{d}$, $s$, $\bar{s}$, and
we use the leading order GRV98 set \cite{Gluck:1998xa} for the parton
distributions, at the scale $\mu^2 = \boldsymbol{K}_{\gamma\perp}^2$. 

Our results for $R$ as a function of  $y$ are shown in Fig.\ \ref{fig:ratio}
at some fixed values of the variables $x_1$, $x_2$ and $|\boldsymbol{K}_{\gamma\perp}|$,
typical of the Tevatron experiments \cite{D0:Atr}.
The values of $x_1$ and $x_2$ considered correspond to their average when both the photon and the jet are  in the 
central rapidity region, where $\eta_j\approx \eta_{\gamma}\approx 0$ and 
$x_1\approx x_2$.
In this case  $y\approx 0.5$, where $R$ turns out to be largest. 
Evidently, 
$R$ increases as $x_1$ and $x_2$ increase, due to the 
small contribution, in the denominator, of the gluon distributions $f_1^g(x)$ 
in the valence region.  

Hence, we see that the asymmetry ${\cal A}$ is a product of a large 
Drell-Yan asymmetry term $\nu$ and a factor $R$ that is estimated to be in the
10\%-50\% range for Tevatron kinematics. This leads us to conclude that an
asymmetry ${\cal A}$ in the order of 5\%-15\%  is possible in the central 
region.  
This could allow a study of the distribution function $h_1^{\perp\,q}$ in 
$p\,\bar p \rightarrow \gamma \,{\rm jet}\, X$ at the Tevatron.

\section{Summary and Conclusions}

In this paper we have calculated the cross section of the  
process $p \,\bar{p} \rightarrow \gamma \,{\rm jet}\, X $ within a 
generalized factorization scheme, taking into account the transverse 
momentum of the partons in the initial proton and antiproton.
In particular, we have studied the contribution from the T-odd, spin
and transverse momentum dependent
parton distribution $h_1^{\perp\,q}$, which leads to an azimuthal asymmetry
similar to the one observed in the Drell-Yan process. Based on the fact that
the latter asymmetry is large in $\pi^- \,N$ scattering and therefore 
most likely also in $p \, \bar{p}$ collisions, we have obtained an estimate 
for the analogous asymmetry in photon-jet production. The latter 
asymmetry is expected to be a factor of 2-10 smaller for typical Tevatron 
kinematics, which may still be sufficiently large to be measurable. 
This would offer a new possibility of measuring T-odd effects using this
high energy collider. A similar measurement could be performed at $p \, p$ 
colliders as well, however, there one expects a significantly smaller 
contribution due to the absence of valence antiquarks. 

The asymmetry measurement itself requires the reconstruction of both the
length and the direction of the photon and jet momenta transverse to the
beam. The angular asymmetry is then a $\cos 2\phi$ asymmetry, with the angle
$\phi$ given by the difference between the angle of either one of the two 
transverse momenta, $\boldsymbol{K}_{j\perp}$ or
$\boldsymbol{K}_{\gamma\perp}$, and the angle of their sum, 
$\boldsymbol{q}_\perp = \boldsymbol{K}_{j\perp}+\boldsymbol{K}_{\gamma\perp}$. 
Realizing the fact that the uncertainty in the latter angle 
$\phi_\perp$ may be rather large
when relatively small $|\boldsymbol{q}_\perp|$ values are considered, 
it may, as an initial step, be convenient 
to integrate the angular distribution over four quadrants which can then be 
added and subtracted in the appropriate way to gain statistics.   
The asymmetry is found to be largest in the central rapidity region,
where $\eta_j\approx \eta_{\gamma} \approx 0$.

\section*{Acknowledgments}
We would like to thank Cedran Bomhof for useful comments and discussions.
This research is part of the
   research program of the ``Stichting voor Fundamenteel Onderzoek der
   Materie (FOM)'', which is financially supported by the ``Nederlandse
   Organisatie voor Wetenschappelijk Onderzoek (NWO)''.


\begin{thebibliography}{99}

\bibitem{Falciano}
  S.~Falciano {\it et al.}  [NA10 Collaboration],
  Z.\ Phys.\ C {\bf 31} (1986) 513.

\bibitem{Guanziroli}
  M.~Guanziroli {\it et al.}  [NA10 Collaboration],
  Z.\ Phys.\ C {\bf 37} (1988) 545.

\bibitem{Conway}
  J.~S.~Conway {\it et al.},
  Phys.\ Rev.\ D {\bf 39} (1989) 92.

\bibitem{Brandenburg-93}
  A.~Brandenburg, O.~Nachtmann and E.~Mirkes,
  Z.\ Phys.\ C {\bf 60} (1993) 697 .

\bibitem{Lam-78}
  C.~S.~Lam and W.~K.~Tung,
  Phys.\ Rev.\ D {\bf 18} (1978) 2447.


\bibitem{Lam-80}
  C.~S.~Lam and W.~K.~Tung,
  Phys.\ Rev.\ D {\bf 21} (1980) 2712.

\bibitem{AL-82}
  E.~N.~Argyres and C.~S.~Lam,
  Phys.\ Rev.\ D {\bf 26} (1982)  114.


\bibitem{Brandenburg-94}
  A.~Brandenburg, S.~J.~Brodsky, V.~V.~Khoze and D.~Muller,
  Phys.\ Rev.\ Lett.\  {\bf 73} (1994) 939.

\bibitem{Eskola-94}
  K.~J.~Eskola, P.~Hoyer, M.~Vanttinen and R.~Vogt,
  Phys.\ Lett.\ B {\bf 333} (1994) 526.

\bibitem{Boer:2002ju}
  D.~Boer, S.~J.~Brodsky and D.~S.~Hwang,
  Phys.\ Rev.\ D {\bf 67} (2003) 054003.

\bibitem{Lu:2004hu}
  Z.~Lu and B.~Q.~Ma,
  Phys.\ Rev.\ D {\bf 70} (2004) 094044.

\bibitem{Boer-04}
  D.~Boer, A.~Brandenburg, O.~Nachtmann and A.~Utermann,
  Eur.\ Phys.\ J.\ C {\bf 40} (2005) 55.

\bibitem{Lu:2005rq}
  Z.~Lu and B.~Q.~Ma,
  Phys.\ Lett.\ B {\bf 615} (2005) 200.


\bibitem{Gamberg:2005ip}
  L.~P.~Gamberg and G.~R.~Goldstein,
  Phys.\ Lett.\  B {\bf 650} (2007)  362.

\bibitem{Brandenburg:2006xu}
  A.~Brandenburg, A.~Ringwald and A.~Utermann,
  Nucl.\ Phys.\  B {\bf 754}  (2006) 107.

\bibitem{Boer:1999mm}
  D.~Boer,
  Phys.\ Rev.\  D {\bf 60}  (1999) 014012.

\bibitem{Zhu:2006gx}
  L.~Y.~Zhu {\it et al.}  [FNAL-E866/NuSea Collaboration],
  Phys.\ Rev.\ Lett.\  {\bf 99} (2007) 082301.


\bibitem{Barone:2006ws}
  V.~Barone, Z.~Lu and B.~Q.~Ma,
  Eur.\ Phys.\ J.\  C {\bf 49} (2007) 967.


\bibitem{Mirkes:1992hu}
  E.~Mirkes,
  Nucl.\ Phys.\ B {\bf 387} (1992) 3.

\bibitem{Mirkes:1994dp}
  E.~Mirkes and J.~Ohnemus,
  Phys.\ Rev.\ D {\bf 51} (1995) 4891.


\bibitem{Acosta:2005dn}
  D.~E.~Acosta {\it et al.}  [CDF Collaboration],
  Phys.\ Rev.\  D {\bf 73} (2006) 052002 .

\bibitem{Bourrely:1994sc}
  C.~Bourrely and J.~Soffer,
  Nucl.\ Phys.\  B {\bf 423} (1994) 329.

\bibitem{Boer:2000er}
  D.~Boer,
  Phys.\ Rev.\  D {\bf 62} (2000) 094029.


\bibitem{Kumar:2007mf}
  A.~Kumar  [D\O\ Collaboration],
  arXiv:0710.0415 [hep-ex].

\bibitem{Pietrycki:2007xr}
   T.~Pietrycki and A.~Szczurek,
   Phys.\ Rev.\  D {\bf 76} (2007) 034003.

\bibitem{Boer:2001he}
  D.~Boer,
  Nucl.\ Phys.\  B {\bf 603} (2001) 195 .

\bibitem{Bacchetta:2007sz}
  A.~Bacchetta, C.~Bomhof, U.~D'Alesio, P.~J.~Mulders and F.~Murgia,
  Phys.\ Rev.\ Lett.\ {\bf 99} (2007) 212002.

\bibitem{Boer:1997nt}
  D.~Boer and P.~J.~Mulders,
  Phys.\ Rev.\  D {\bf 57} (1998) 5780.


\bibitem{Mulders:2000sh}
  P.~J.~Mulders and J.~Rodrigues,
  Phys.\ Rev.\  D {\bf 63} (2001) 094021.

\bibitem{Bomhof:2006ra}
  C.~J.~Bomhof and P.~J.~Mulders,
  JHEP {\bf 0702} (2007) 029.

\bibitem{Bomhof:2007xt}
  C.~J.~Bomhof and P.~J.~Mulders,
  arXiv:0709.1390 [hep-ph],
  to be publ.\ Nucl.\ Phys.\ B.

\bibitem{Collins:2002kn}
  J.~C.~Collins,
  Phys.\ Lett.\  B {\bf 536} (2002) 43.

\bibitem{Belitsky:2002sm}
  A.~V.~Belitsky, X.~Ji and F.~Yuan,
  Nucl.\ Phys.\  B {\bf 656} (2003) 165.

\bibitem{Boer:2003cm}
  D.~Boer, P.~J.~Mulders and F.~Pijlman,
  Nucl.\ Phys.\  B {\bf 667} (2003) 201.


\bibitem{Collins:2007nk}
   J.~Collins and J.~W.~Qiu,
   Phys.\ Rev.\  D {\bf 75} (2007) 114014.

\bibitem{Vogelsang:2007jk}
   W.~Vogelsang and F.~Yuan,
   Phys.\ Rev.\  D {\bf 76} (2007) 094013.

\bibitem{Collins:2007jp}
   J.~Collins,
   arXiv:0708.4410 [hep-ph].

\bibitem{Boer:2006eq}
   D.~Boer and W.~Vogelsang,
   Phys.\ Rev.\  D {\bf 74} (2006) 014004.

\bibitem{Berger:2007jw}
   E.~L.~Berger, J.~W.~Qiu and R.~A.~Rodriguez-Pedraza,
   Phys.\ Rev.\  D {\bf 76} (2007) 074006.


\bibitem{D0:Atr}
 O.~Atramentov, talk given at the XV International Workshop on Deep-Inelastic Scattering and Related Subjects (DIS2007), Munich, Germany, 16-20 April 2007. 



\bibitem{Mulders:1995dh}
  P.~J.~Mulders and R.~D.~Tangerman,
  Nucl.\ Phys.\  B {\bf 461} (1996) 197
  [Erratum-ibid.\  B {\bf 484} (1997) 538].


\bibitem{D'Alesio:2004up}
  U.~D'Alesio and F.~Murgia,
  Phys.\ Rev.\  D {\bf 70} (2004) 074009.
\bibitem{Gluck:1998xa}
  M.~Gl\"uck, E.~Reya and A.~Vogt,
  Eur.\ Phys.\ J.\  C {\bf 5} (1998) 461. 

\end{thebibliography}
\end{document}